\documentstyle[12pt]{article}
\newcommand{\ltwid}{\raise.3ex\hbox{$<$\kern-.75em\lower1ex\hbox{$\sim
$}}}

\newcommand{\td}{\nabla}
\begin{document}
\vspace{10mm}
\begin{center}
{\bf CALCULATION OF DE WITT--SCELEY--GILKEY COEFFICIENTS FOR MINIMAL
FOURTH--ORDER OPERATOR}
\end{center}
\vspace{10mm}
\begin{center}
{\sl I.A.Shovkovy}\\
{\sl Bogolyubov Institute for Theoretical Physics},\\
{\sl 252143 Kiev, Ukraine}
\end{center}

\vfill
\begin{abstract}
De Witt--Sceley--Gilkey coefficients  are  calculated  for the most
general minimal differential fourth--order operator on Riemannian space of
an arbitrary dimension.
\end{abstract}

\vfill
\eject

As is well known, the diagonal matrix elements of the heat kernel,
$\langle x|\exp(-tA)|x\rangle$ where $A$ is an elliptic  differential
operator, can be expanded into asimptotic series at $t\to \infty$ [1].
These series have the form:
\begin{equation}
\langle x|\exp(-tA)|x\rangle=\sum_{m=0}^{\infty} E_m(x|A) t^{(m-n)/2r},
\end{equation}
where $2r$ is the operator order and coefficients $E_m(x|A)$, called
De Witt--Seeley--Gilkey (DWSG) coefficients (sometimes they called by names
of Schwin\-ger, Ha\-da\-mard and Minakshisundaram), depend on the coefficient
functions of the operator $A$ and their derivatives.

There are several methods for calculating the asymptotic expansion (1).
In this paper we use the method developed in [2] which is
based on the theory of pseudodifferential operators.
The advantage of this method is its explicit covariance with respect to
general coordinate transformations.

We consider the fourth--order minimal operator of the form:
\begin{equation}
\Delta=\Box^2 +B^{\mu \nu \lambda}\td_{\mu }\td_{\nu }\td_{\lambda}+V^{\mu \nu}
\td_{\mu}\td_{\nu}+N^{\mu}\td_{\mu}+X,
\end{equation}
where $\td_{\mu}$ is a covariant derivative, including both affine
and spinor connection (torsion is absent), coefficient functions
$B^{\mu\nu\lambda}$, $V^{\mu\nu}$, $N^{\mu}$ and $X$ are matrices.
Such operators are used in quantum gravity with quadratic curvature
term in lagrangian [3].
In paper [2] practically the same problem was posed.
But, in order to simplify calculations, the term with $B^{\mu\nu\lambda}$
was omitted there. Below we shall present those additional terms in DWSG
coefficients for operator (2) that contain tensor $B^{\mu\nu\lambda}$.

We suppose that the tensor $B^{\mu\nu\lambda}$ is symmetric in all
indices.

The equation for the amplitude $\sigma(x,x',k;\lambda)$ (for the formalism
used here see [2]) is:
\begin{eqnarray}
&&\bigg[(\td_{\mu}+i\td_{\mu}l)(\td^{\mu}+i\td^{\mu}l)(\td_{\nu}+i\td_{\nu}l)
(\td^{\nu}+i\td^{\nu}l)+ \nonumber \\
&&+B^{\mu \nu \lambda}(\td_{\mu}+i\td_{\mu}l)
(\td_{\nu}+i\td_{\nu}l)(\td_{\lambda}+i\td_{\lambda}l)+\nonumber \\
&&+V^{\mu \nu}(\td_{\mu}+i\td_{\mu}l)(\td_{\nu}+i\td_{\nu}l)
+N^{\mu}(\td_{\mu}+i\td_{\mu}l)+\nonumber \\
&&+X-\lambda\bigg]
\sigma(x,x',k;\lambda)=I(x,x').
\end{eqnarray}
Following the paper [2], we seek for the solution to the equation
(3) in the form of series over an auxilary parameter $\epsilon$ :
$\sigma=\sum_{m=0}^{\infty}\epsilon^{4+m}\sigma_m$. After changing
$l\to l/\epsilon$ and $\lambda\to \lambda/\epsilon^4$ in the equation, we
obtain the following recursion expressions:
\begin{eqnarray}
((\td^{\mu}l\td_{\mu}l)^2-\lambda)\sigma_m+A_1\sigma_{m-1}+
A_2\sigma_{m-2}&+& \nonumber\\
+A_3\sigma_{m-3}+A_4\sigma_{m-4}&=&0 ,
\end{eqnarray}
where $\sigma_{m}\equiv 0$ for $m<0$ by definition and
\begin{eqnarray*}
A_1&=&-i\bigg(2\Box l\td^{\mu}l \td_{\mu}l
+4\td^{\mu}l \td^{\nu}l \td_{\mu}\td_{\nu}l+\\
&&+4\td^{\mu}l \td_{\mu}l \td^{\nu}l \td_{\nu}+B^{\mu \nu \lambda}
\td_{\mu}l \td_{\nu}l \td_{\lambda}l\bigg),
\end{eqnarray*}
\begin{eqnarray*}
A_2&=&-\bigg((\Box l)^2 +2\td^{\mu} \td^{\nu}l \td_{\mu} \td_{\nu}l
+2 \td^{\mu}l(\td_{\mu}\Box l+\Box\td_{\mu}l)+ \\
&&+3B^{\mu \nu \lambda}\td_{\mu} \td_{\lambda}l\td_{\nu}l
+V^{\mu\nu}\td_{\mu}l\td_{\nu}l+2\td^{\mu}l \td_{\mu}l\Box+\\
&&+4\td^{\mu}l\td^{\nu}l\td_{\mu}\td_{\nu}
+\left(4\Box l \td^{\mu}l+4\td_{\nu}l\td^{\mu}\td^{\nu}l+\right.\\
&&+\left. 4\td_{\nu}l\td^{\nu}\td^{\mu}l+3B^{\mu\nu\lambda}
\td_{\nu}l\td_{\lambda}l\right)\td_{\mu}
\bigg),
\end{eqnarray*}
\begin{eqnarray*}
A_3&=&i\bigg(\Box\Box l+B^{\mu \nu \lambda}\td_{\mu}\td_{\nu}\td_{\lambda}l
+V^{\mu\nu}\td_{\mu}\td_{\nu}l+N^{\mu}\td_{\mu}l+  \\
&&+\left(2\td^{\mu}\Box l +2\Box\td^{\mu}l+3B^{\mu \nu \lambda}
\td_{\nu} \td_{\lambda}l+2V^{(\mu\nu)}\td_{\nu}l\right)\td_{\mu}+\\
&+&2\Box l\Box+\left(4\td^{\mu}\td^{\nu}l+3B^{\mu\nu\lambda}\td_{\lambda}l
\right)\td_{\mu}\td_{\nu}+ \\
&&+2\td^{\mu}l\left(\td_{\mu}\Box+\Box\td_{\mu}\right)
\bigg),\nonumber\\
A_4&=&\Delta.
\end{eqnarray*}

Using technique of the paper [2], the coefficients $[\sigma_m]$
($m=0,1,2,3,4$) can be easily found. To avoid unnecessary complications in
this paper we do not write down them (for details see [4]).
This functions lead to the following expressions for DWSG coefficients:
\begin{eqnarray*}
E_2(x|\Delta)&=&\frac{1}{(4\pi)^{n/2}}\frac{\Gamma((n-2)/4)}
{2\Gamma((n-2)/2)}\bigg(\frac{1}{6}R+\frac{1}{2n}V_{\alpha}^{\;\alpha}
-\frac{3}{4n}\td_{\alpha}b^{\alpha}-\nonumber\\
&&-\frac{3}{16n(n+4)}\left(3b_{\beta}
b^{\beta}+2B^{\alpha\beta\gamma}B_{\alpha\beta\gamma}\right)\bigg),\\
E_4(x|\Delta)&=&\frac{1}{(4\pi)^{n/2}}\frac{\Gamma((n+4)/4)}
{\Gamma((n+2)/2)}\bigg(h_0+h_1+h_2+h_3\bigg),
\end{eqnarray*}
where
\begin{eqnarray*}
h_1&=&\frac{1}{4}\td_{\alpha}B^{\alpha\beta\gamma}R_{\beta\gamma}-
\frac{1}{8}\td_{\alpha}b^{\alpha}R-\frac{1}{4}b_{\beta}\td_{\alpha}
W^{\alpha\beta}+\\
&+&\frac{n}{4(n+2)}\td_{\alpha}\td_{\beta}\td_{\gamma}B^{\alpha\beta\gamma}
-\frac{3}{4(n+2)}\left(-\frac{1}{2}\{b_{\alpha},N^{\alpha}\}\right.+\\
&+&\left. V^{\alpha\beta}\td^{\gamma}B_{\alpha\beta\gamma}
+b_{\alpha}\td_{\beta}V^{(\alpha\beta)}+\frac{1}{2}V_{\beta}^{\;\beta}
\td_{\alpha}b^{\alpha}\right)+\\
&+&\frac{n+4}{12(n+2)}\td_{\beta}(b_{\alpha}R^{\alpha\beta}+
3b_{\alpha}W^{\beta\alpha})-\\
&-&\frac{n+4}{8(n+2)}\{\td^{\alpha}\td_{\beta}\td_{\alpha}\}
b^{\beta}+\frac{n+1}{12(n+2)}[b^{\alpha},\td^{\beta}W_{\beta\alpha}]+\\
&+&\frac{1}{4(n+2)}\left([V^{(\alpha\beta)},\td_{\alpha}b_{\beta}]
+[B^{\alpha\beta\gamma},\td_{\alpha}V_{\beta\gamma}]-\right.\\
&-&\left. [\td_{\alpha}V^{(\alpha\beta)},b_{\beta}]
-[\td_{\alpha}B^{\alpha\beta\gamma},V_{\beta\gamma}]\right)+\\
&+&\frac{1}{8(n+2)}\left([b^{\alpha},
\td_{\alpha}V_{\beta}^{\;\beta}]-[\td_{\alpha}b^{\alpha},
V_{\beta}^{\;\beta}]\right),
\end{eqnarray*}
\begin{eqnarray*}
h_2&=&-\frac{\{(B_{\circ\circ\circ})^2,V_{\circ\circ}\}+
B_{\circ\circ\circ}V_{\circ\circ}B_{\circ\circ\circ}}{3\cdot 2^5(n+6)(n+2)}-
\frac{R(B_{\circ\circ\circ})^2}{96(n+2)}-\\
&-&\frac{n+8}{96(n+6)(n+2)}
\bigg(\Box(B_{\circ\circ\circ})^2-\td_{\alpha}B_{\circ\circ\circ}
\td^{\alpha}B_{\circ\circ\circ}+\\
&+&18B^{\alpha}_{\;\circ\circ}
\td_{(\alpha}\td_{\beta)}B^{\beta}_{\;\circ\circ}\bigg)
+\frac{3(n+4)}{16(n+6)(n+2)}\bigg(\td_{\alpha}\td_{\beta}b_{\gamma}
B^{\alpha\beta\gamma}+\\
&+&\td_{\alpha}\td_{\beta}(B^{\alpha\beta\gamma}b_{\gamma})+
\td_{(\alpha}(\td_{\beta)}B^{\alpha}_{\;\circ\circ}B^{\beta}_{\;\circ\circ})
\bigg)+\\
&+&\frac{3(3n+16)}{16(n+6)(n+2)}
\bigg(b_{\alpha}\td_{\beta}\td_{\gamma}B^{\alpha\beta\gamma}+
[\td_{\alpha}b_{\gamma},\td_{\beta}B^{\alpha\beta\gamma}]
\bigg)+\\
&+&\!\frac{3}{16(n+2)}\bigg(\!\td_{\alpha}(B^{\alpha}_{\;\circ\circ}
\td_{\beta}B^{\beta}_{\;\circ\circ}\!-\!B^{\beta}_{\;\circ\circ}\td_{\beta}
B^{\alpha}_{\;\circ\circ})\!-\!B^{\alpha}_{\;\circ\circ}W_{\alpha\beta}
B^{\beta}_{\;\circ\circ}\!+\\
&+&2B^{\alpha}_{\;\circ\circ}B^{\beta}_{\;\circ\circ}W_{\alpha\beta}
+B^{\alpha}_{\;\circ\circ}B^{\beta}_{\;\circ\circ}
R_{\alpha\beta}-4B^{\alpha\beta\sigma}B_{\sigma}^{\;\gamma\delta}
R_{\alpha\gamma\beta\delta}\bigg)+\\
&+&\frac{3R_{\alpha\beta}}{16(n+2)}
\{B^{\alpha\beta\gamma},b_{\gamma}\},
\end{eqnarray*}
\begin{eqnarray*}
h_3&=&\frac{(B_{\circ\circ\circ})^4}{3\cdot 2^9(n+10)(n+6)(n+2)}+\\
&+&\frac{2\td_{\circ}(B_{\circ\circ\circ})^3+[(B_{\circ\circ\circ})^2,
\td_{\circ}B_{\circ\circ\circ}]}{3\cdot 2^7(n+6)(n+2)}-\\
&-&\frac{\{B_{\alpha\circ\circ},B_{\circ\circ\circ}\}\td^{\alpha}
B_{\circ\circ\circ}+B_{\alpha\circ\circ}\td^{\alpha}B_{\circ\circ\circ}
B_{\circ\circ\circ}}{32(n+6)(n+2)}.
\end{eqnarray*}

Since the expression for $h_0$ does not include terms with
$B^{\mu\nu\lambda}$ (and, consequently, coincides with $h_0$ found in
[2]), it is not presented here.

As for the notations, all of them coincide with those in [2],
except for the following: $b_{\alpha}=B_{\alpha\beta}^{\;\;\;\beta}$ and
$T_{\circ\circ\cdots\circ}=g^{\{\alpha\beta\ldots\lambda\}}
T_{\alpha\beta\ldots\lambda}$.

DWSG coefficients for the operator (2) on a manifold of dimension $n=4$
were found in [5,6]. In order to check the result
obtained here we presented the expressions in the same form as in
[5].

Substituting $n=4$ in $E_2(x|\Delta)$ and $E_4(x|\Delta)$, one can see that,
except for three places, our result is the same.

The indicated differences are the following:

{\em a)} the expression for $h_2$, obtained here, contains two additional
terms:
\begin{eqnarray*}
&\frac{3}{16(n+2)}\left(B^{\alpha}_{\;\circ\circ}B^{\beta}_{\;\circ\circ}
R_{\alpha\beta}-4B^{\alpha\beta\sigma}B_{\sigma}^{\;\gamma\delta}
R_{\alpha\gamma\beta\delta}\right),&
\end{eqnarray*}
which are absent in the above mentioned paper. However, these terms are
present in the second paper [6], where the result has slightly
different form. We can conclude that in  [5] there is a mistake.

{\em b)} in the expression for $h_1$, instead of the term
\begin{eqnarray*}
&(1/4)\td_{\alpha}B^{\alpha\beta\gamma}R_{\beta\gamma},&
\end{eqnarray*}
in the mentioned paper we find
\begin{eqnarray*}
&(1/4)\td_{\alpha}B^{\alpha\beta\gamma}\td_{\gamma}\td_{\beta}R.&
\end{eqnarray*}
As in the first case comparison with [6] solves the question
again. So we conclud that in [5] there is one more misprint.

{\em c)} obtained here expression for $h_1$ contains one more term which is
absent in [5]:
\begin{equation}
-\frac{1}{8(n+2)}\left[\td_{\alpha}b^{\alpha},V\right].
\end{equation}
Unfortunately, in this case we cannot solve the question so easy as in the
previous two cases. As for the paper [6], it could not help
us now because the authors limited themselves to the operators with
commuting coefficient functions $B^{\mu\nu\lambda}$ and $V^{\mu\nu}$ and,
consequently, all expressions like (5) are identically equal to zero.

\begin{center}
{\Large\bf References}
\end{center}
\begin{enumerate}
\item Seeley R.T. Complex powers of an elliptic operator.//Proc. Symp. Pure
Math., Amer. Math. Soc.-1967.- {\bf 10}.-p.288-307.
\item  Gusynin V.P. Seeley--Gilkey coefficients for fourth--order operators on
a
riemannian manifold.// Nucl. Phys.-1990.-{\bf B333},N2.-p.296-316.
\item  Fradkin E.S., Tseytlin A.A. Renormalizable asymptotically free quantum
theory of gravity.// Nucl. Phys.-1982.-{\bf B201},N3.-p.469-491.
\item  Shovkovy I.A. Calculation of De Witt--Seeley--Gilkey coefficients for
the
general minimal fourth--order operator.-Kiev, 1992.-12p.-(Preprint
/ Bogolyubov Inst. for Theor. Phys.; ITP-92-19P).
\item  Lee H.W., Pak P.Y. and Shin H.K. New algorithm for asymptotic expansions
of
the heat kernel.// Phys. Rev.-1987.-{\bf D35},N8.-p.2440-2447.
\item  Barvinsky A.O., Vilkovisky G.A. The generalized Schwinger--De Witt
technique in gauge theories and quantum gravity.// Phys. Rep.-1985.-
{\bf 119},N1.-p.1-74.
\end{enumerate}
\end{document}